\documentclass[aps,pra,preprint,groupedaddress]{revtex4}
\usepackage{epsfig,subfigure,subfloat,graphicx,amssymb,latexsym,amsthm
,amsmath,natbib,slashed,dsfont,color,nicefrac}
\usepackage{color}

\newcommand{\beq}{\begin{equation}}
\newcommand{\eeq}{\end{equation}}
\newcommand{\beqa}{\begin{eqnarray}}
\newcommand{\eeqa}{\end{eqnarray}}

\newcommand{\ben}{\begin{enumerate}}
\newcommand{\een}{\end{enumerate}}
\newcommand{\bit}{\begin{itemize}}
\newcommand{\eit}{\end{itemize}}
\newcommand{\bpm}{\begin{pmatrix}}
\newcommand{\epm}{\end{pmatrix}}

\newcommand{\pad}{\partial}
\newcommand{\pd}[2]{\frac{\partial #1}{\partial #2}}
\newcommand{\pdn}[3]{\frac{ {\partial}^{#1} #2}{{\partial} #3 ^{#1}}}

\def\N{{\cal N}}
\def\H{{\cal H}}

\def\lt{\left}
\def\rt{\right}

\def\bra#1{{\langle #1|}}
\def\ket#1{{| #1\rangle}}

\def\intl#1#2{\int\limits_{#1}^{#2}}
\def\bracket#1#2 {\mathinner{\langle{#1}|{#2}\rangle}}

\def \Dl {\Delta}
\def \dl {\delta}

\def \sg {\sigma}

\def \nb {\nabla}

\def\->{\rightarrow}

\def\br{{\bs r}}
\def\bk{{\bs k}}

\def\bq{{\bs q}}
\def\bp{{\bs p}}

\def \bs {\boldsymbol}

\def \. {\cdot}

\def \. {\cdot}

\begin{document}

%\preprint{}

\title{Breakdown of Dirac Dynamics in Honeycomb Lattices 
due to Nonlinear Interactions}


\author{Omri Bahat-Treidel}
\affiliation{Department of Physics, Technion-Israel Institute of
Technology, Technion City, Haifa 32000, Israel}
\author{Or Peleg}
\affiliation{Department of Physics, Technion-Israel Institute of
Technology, Technion City, Haifa 32000, Israel}
\author{Hrvoje Buljan}
\affiliation{Department of Physics, University of Zagreb, PP 332, 10000 Zagreb, Croatia}
\author{Mordechai Segev}
\affiliation{Department of Physics, Technion-Israel Institute of
Technology, Technion City, Haifa 32000, Israel}
%\email[]{msegev@tx.technion.ac.il}
%\thanks{}
\affiliation{Department of Physics, Technion-Israel Institute of
Technology, Technion City, Haifa 32000, Israel}


\today

\begin{abstract}  % abstract
We study the dynamics of coherent 
waves in nonlinear honeycomb lattices 
and show that nonlinearity breaks down the Dirac dynamics. As an example, we 
demonstrate that even a weak nonlinearity has major qualitative effects one of the hallmarks of honeycomb lattices: conical diffraction. 
Under linear conditions, a circular input wave-packet associated with the Dirac point 
evolves into a ring, but   
even a weak nonlinearity alters the evolution such that the emerging 
beam possesses triangular symmetry, and populates Bloch modes outside of the 
Dirac cone region. Our results are presented in the context of optics, 
but we propose a scheme to observe equivalent phenomena in 
Bose-Einstein condensates. 
\end{abstract}

%\keywords{Klein paradox}

\maketitle

%\section{}
%\subsection{}
%\subsubsection{}

%%%%%%%%%%%%%% body of the paper %%%%%%%%%%%%%%%%%%%%%%
The past decade has witnessed considerable interest in honeycomb lattices in many fields, 
ranging from carbon nanotubes \cite{CN} and graphene \cite{PhysRevB.65.245420, 
Novoselov_2005, PhysRevLett.95.146801} in condensed matter, cold atoms in honeycomb 
optical lattices \cite{PhysRevLett.99.070401, zhu-2007-98, PhysRevB.77.235107, 
PhysRevA.80.043411}, and electromagnetic waves in honeycomb photonic crystals 
\cite{PhysRevLett.100.013904, PhysRevA.75.063813, PhysRevB.78.045122, PhysRevB.80.155103} and photonic 
lattices (waveguide arrays) \cite{PhysRevLett.98.103901,Bahat-Treidel:08, PhysRevLett.104.063901}. 
The unique features of the honeycomb lattice result from its unusual band structure, 
displaying two intersecting bands. The vicinity of the intersection points is 
conical, and excitations residing at the close vicinity of the intersection 
points obey the massless Dirac equation: a relativistic wave equation for 
massless spin-half particles. As a consequence of this unusual linear dispersion 
and the additional degrees of freedom (two sites in each unit cell, referred as 
pseudo-spin), a variety of exciting phenomena have been obtained:  room temperature 
quantum Hall effect in graphene \cite{amersouda:Novoselov2007}, conical diffraction 
in honeycomb photonic lattices \cite{PhysRevLett.98.103901}, negative magnetoresistance 
which implies anti-localization \cite{PhysRevLett.97.016801, PhysRevLett.97.146805, 
PhysRevLett.103.226801}, Klein tunneling \cite{Ando_1998, Katsnelson_2006, kim_2009, 
PhysRevLett.104.063901}, and Zitterbewegung \cite{Katsnelson_Zitterbewegung, 
PhysRevB.80.045416} to name a few. The dynamics of Dirac-like excitations in 
honeycomb lattices has been well studied when the propagation equation is linear 
(interaction-free). However, the dynamics can also be nonlinear, as happens in the 
optical domain due to light-matter interactions, and for atomic Bose-Einstein condensates 
(BEC) due to pair-wise scattering. In either of these, the nonlinear dynamics 
has received little attention in the context of honeycomb lattices. 
The first study of nonlinear dynamics in honeycomb lattices was conducted in \cite{PhysRevLett.98.103901}, demonstrating gap solitons, which had no overlap with Bloch modes residing at the vicinity to the Dirac points. Later studies of nonlinear dynamics in honeycomb lattices were a    
generalization of Dirac approximation \cite{PhysRevA.79.053830, nonlinear_dirac}, 
where a nonlinear version of the massless Dirac equation have been studied.

Here, we study the nonlinear dynamics of waves in honeycomb lattices, 
{where initial wave-packets are} comprised of Bloch waves from the vicinity of the Dirac points. 
Such wave-packets are very well described by the massless Dirac equation. However, 
even the presence of a fairly weak nonlinearity drives the waves away from the 
vicinity of the Dirac points, hence the nonlinearity breaks the Dirac dynamics 
associated with honeycomb lattices. 
We demonstrate this dramatic change in the dynamics by investigating conical 
diffraction: an initial wave-packet with circular symmetry that evolves into two 
concentric rings (separated by a dark ring). Surprisingly, under slightly
nonlinear conditions, the same (circular) input beam evolves into a triangular-ring beam. 
Interestingly, we find that the resulting triangle is rotated by $\pi$ when the 
nonlinearity changes sign, i.e., a given input beam 
subjected to a focusing nonlinearity (attractive interactions for BEC) evolves 
into a triangle that is rotated by $\pi$ with respect 
to the triangle obtained when the same input beam is subjected 
to a defocusing (repulsive) nonlinearity. 
Hence, the presence of nonlinearity (interactions) 
changes the wave dynamics in honeycomb lattices dramatically, introducing important changes to the 
unique phenomena associated with the effective Dirac equation, such as conical 
diffraction and Klein tunneling. Moreover, the nonlinear breakdown the Dirac dynamics - in itself gives rise to interesting new effects.
Our results are presented in the context of optics, however, 
we propose a scheme to observe equivalent phenomena in Bose-Einstein condensates.

\begin{figure}[]
    \center
    {\includegraphics[width=0.215\textwidth]{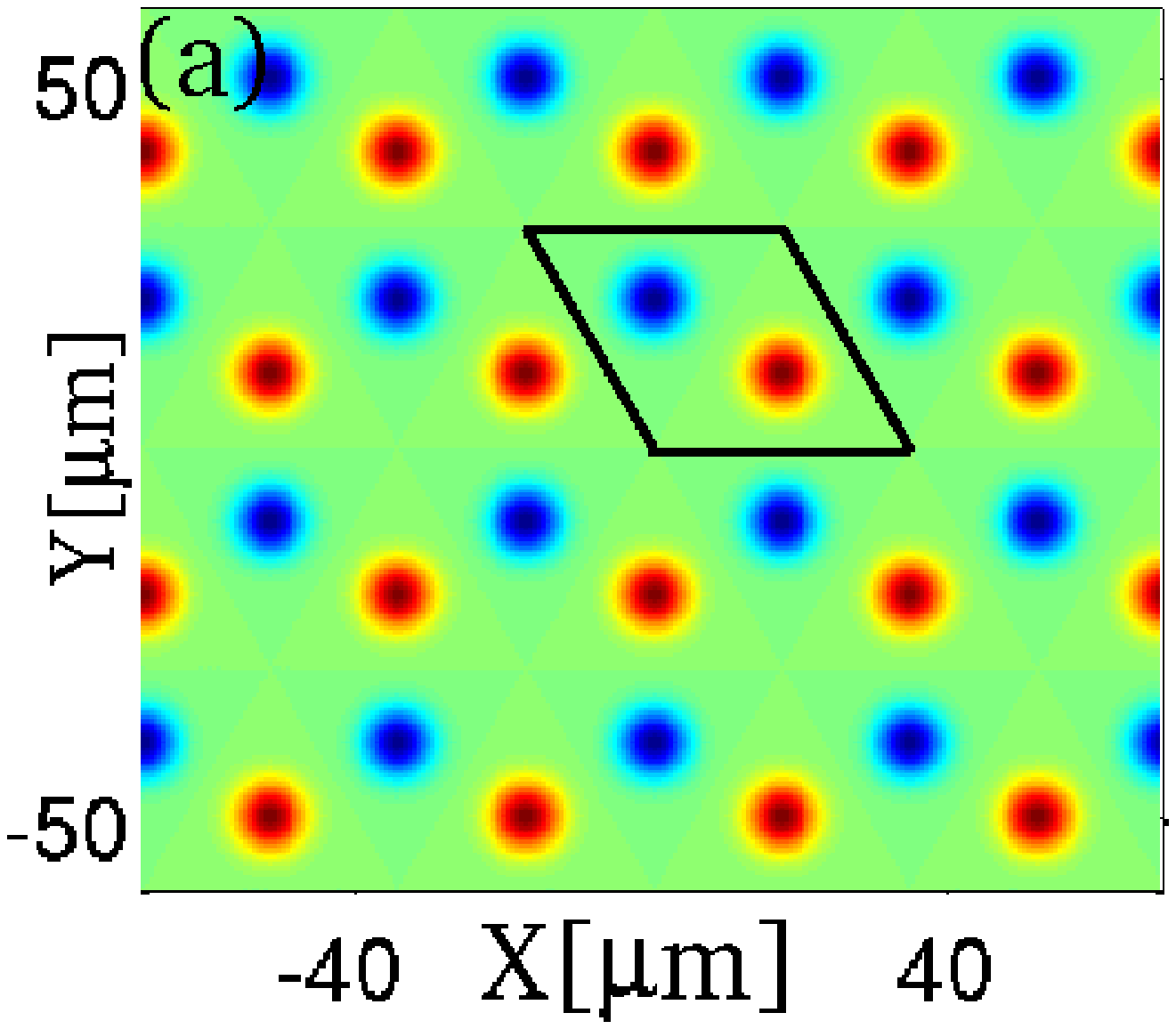}
    \includegraphics[width=0.225\textwidth]{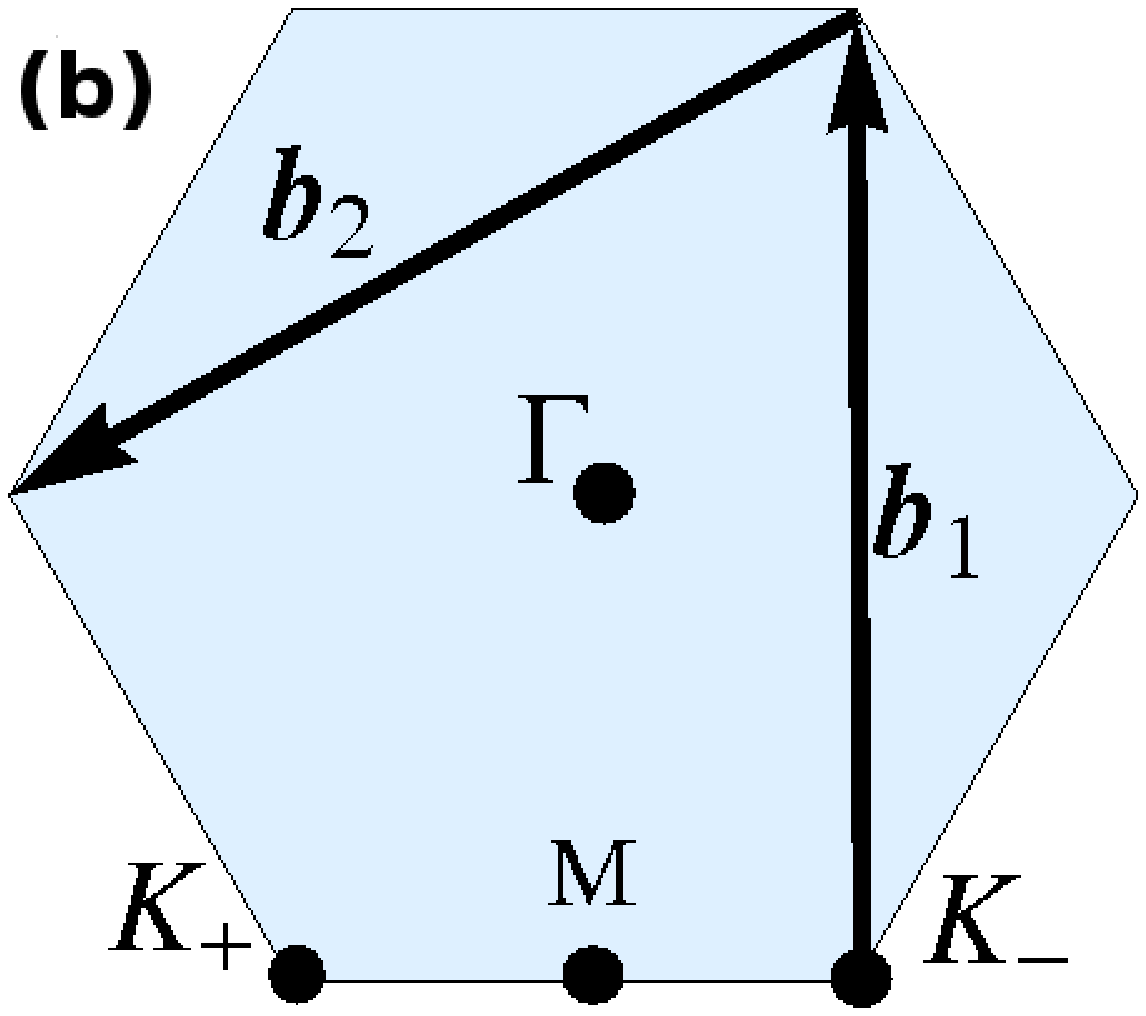}
    \includegraphics[width=0.26\textwidth]{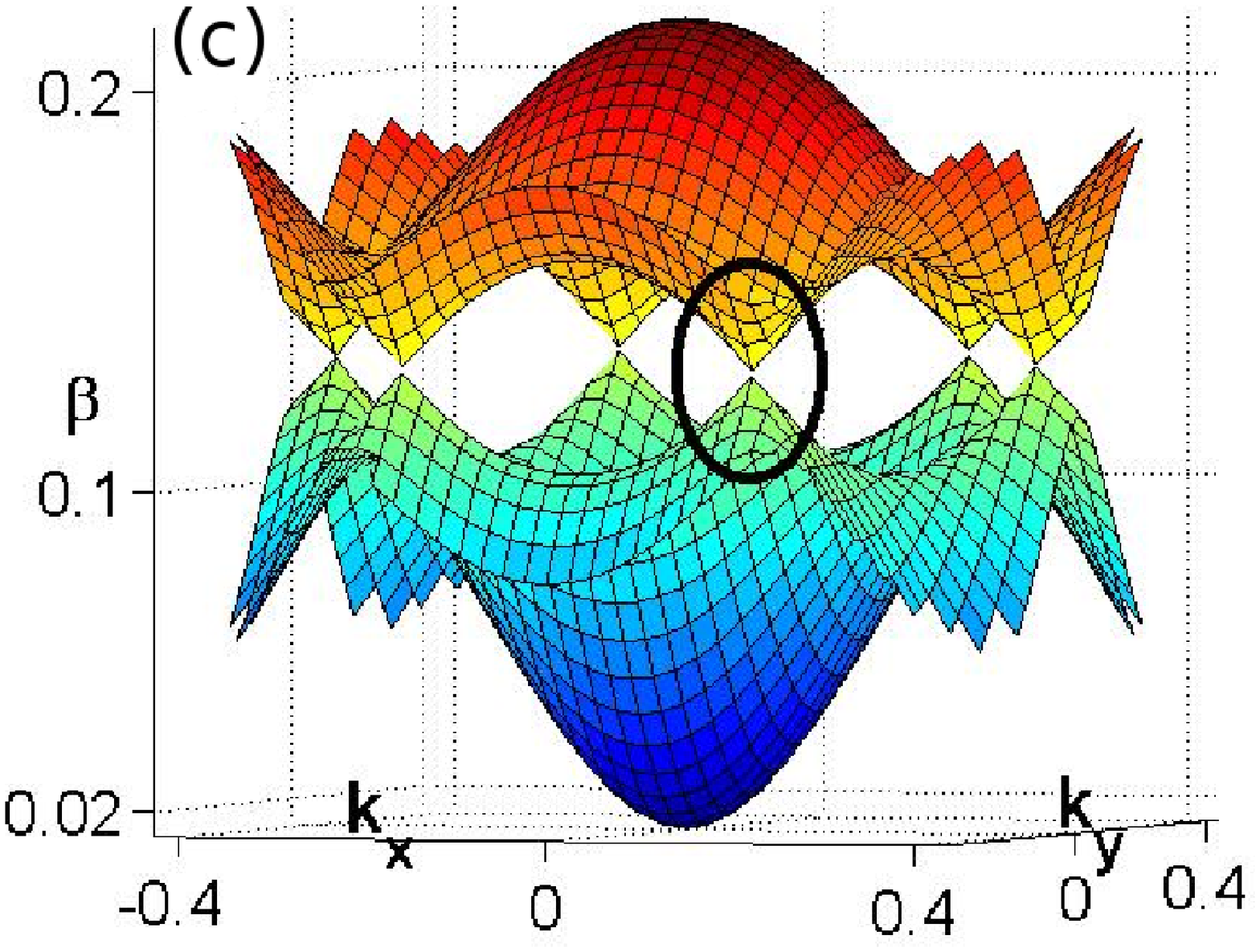}
	\includegraphics[width=0.26\textwidth]{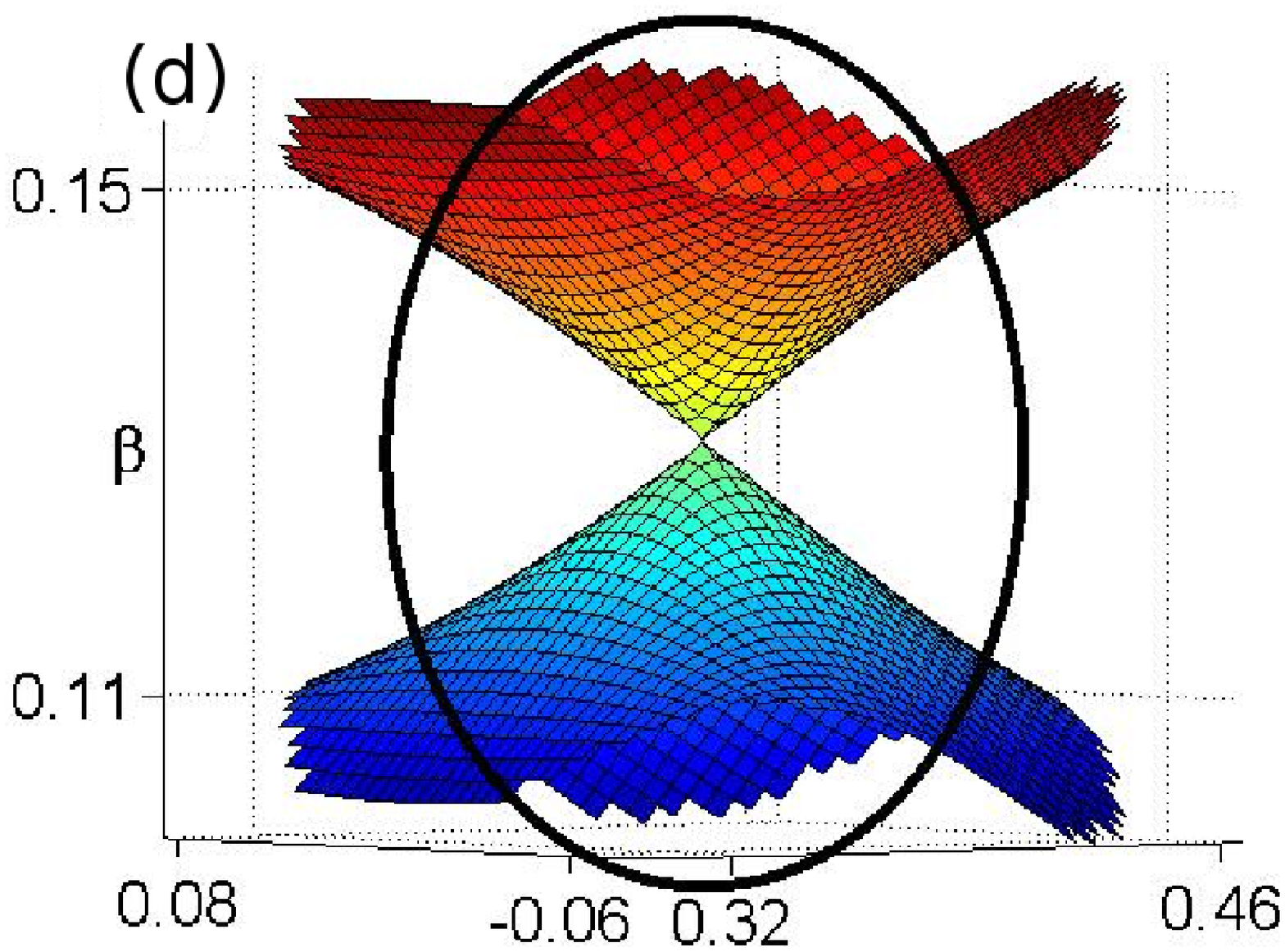}    
    }%
    \caption {(a) Perfect honeycomb lattice that has two sites in a unit cell.
    (b) The first Brillouin zone with the high symmetry points. (c) The first two bands, and (d)
    The vicinity of the intersection (Dirac) point.}
    \label{lat_band}
\end{figure}

The paraxial propagation of a monochromatic field envelope $\psi$ inside a photonic lattice with Kerr non-linearity is described by 
\beq
    i \pdn{}{\psi}{z} = -\frac{1}{2k}\nb^2_{\perp}\psi -
    \frac{k\dl n(x,y)}{n_0} \psi - \frac{k}{n_0}n_2 |\psi|^2 \psi,
    \label{nls}
\eeq %
where $\dl n(x,y)$ is the modulation in the refractive index defining the lattice as shown in Fig. \ref{lat_band} (a), $k$ is the wave-number in the medium, 
$n_0$ the background refractive index, and $n_2$ is the Kerr coefficient. 
The sign of $n_2$ determines the type of nonlinearity, where $n_2 > 0$ 
corresponds to a focusing nonlinearity (attractive interactions in the context of BEC). 
The term $k \dl n/n0$ is referred to as the optical potential.
    
At low intensities ($n_2 |\psi|^2 \ll \max\{\dl n(x,y)\} $), the nonlinear term is negligible and the propagation can be described by linear methods. In the absence of the nonlinear term, Eq.(\ref{nls}) has solutions of the form 
$\psi(x,y,z) = U(x,y) \exp(i \beta z)$ where $U(x,y)$ is a solution of the eigenvalue equation
\beq
	\hat H_0 U(x,y) = \beta U(x,y),\quad \text{where} \quad 
	\hat H_0 \equiv -\frac{1}{2k}\nb^2_{\perp} - \frac{k}{n_0}\dl n(x,y), \label{eigenvalue equation}
\eeq
and the eigenvalue, $\beta$, is the propagation constant (analogous to the energy in quantum mechanics with opposite sign, i.e., $\beta \leftrightarrow -E$). Since $\dl n(x,y)$ is a periodic function in $x$ and $y$, the eigenfunctions of $\hat H_0$ can be chosen to be Bloch waves
\beq
	B_{\bq,m}(\br) = u_{\bq,m}(\br)\exp{(i \bq \cdot \br)}, \label{bloch wave} 
\eeq
where $u_{\bq,n}(x,y)$ has the same periodicity as the potential, $\bq = (q_x,q_y)$ is the lattice momentum, 
$\br = (x,y)$ is the transverse coordinate, and $m$ is the band index. We solve numerically 
Eq.(\ref{eigenvalue equation}) in the first Brillion zone (BZ) (Fig. \ref{lat_band} (b)) and obtain the band structure $\beta(\bq)$ (Fig. \ref{lat_band}(c)). 
Note the intersection of the bands at the corners of the first BZ. 
According to Fig. \ref{lat_band}(a), it appears as if there are six intersection points; 
however, these six points are in fact two triplets of two non-equivalent points 
denoted by $K_{\pm}$ (Fig. \ref{lat_band} (b)). That is, there are only two non-equivalent intersection points, 
since the corners are connected by reciprocal lattice vectors (Fig. \ref{lat_band}(b)). 
The close vicinity of the intersection points is conical (Fig. \ref{lat_band}(d)). 
In the tight binding approximation, the effective Hamiltonian describing the 
Bloch modes from the vicinity of these points is the Dirac Hamiltonian for 
massless particles \cite{PhysRevLett.53.2449}. 
Hence, these points are often referred to as {\it Dirac points}, and the 
band structure in the vicinity of these points forms {\it Dirac cones}. 
We emphasize that Bloch modes that reside outside the Dirac cone cannot be described 
by the Dirac equation, since the propagation constant $\beta(\bq)$ is no longer 
linear in $\bq$, and moreover it does not possess circular symmetry but rather a 
three-fold symmetry (as will be discussed later).

\begin{figure*}[t]
    \center
    {\includegraphics[width=0.96\textwidth]{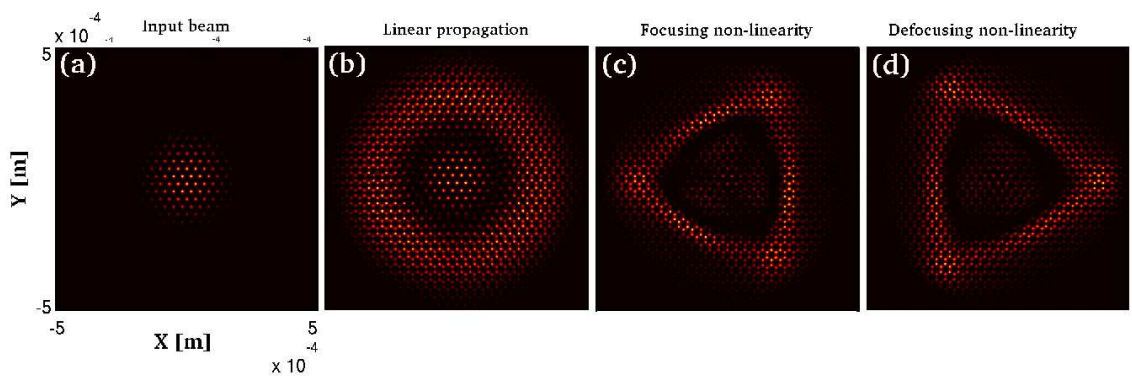}
    }%
    \caption { (a) Input beam from the vicinity of $K_+$. (b) The output of the linear propagation,
    (c) the output of the nonlinear propagation with focusing nonlinearity, 
    and (d) the output of the nonlinear propagation with defocusing nonlinearity.
}
    \label{real_space}
\end{figure*}

Our main interest in the article is the effect of nonlinearity on the extremely unique diffraction pattern, named
{\it conical diffraction,} obtained for an input beam comprised of Bloch modes residing 
in the Dirac cone. Conical diffraction refers to the fact that 
at a large enough propagation distance, the circular (envelope) 
input beam evolves into two bright rings of constant width separated by a dark ring, 
and a vanishing intensity at the origin \cite{PhysRevLett.98.103901, Bahat-Treidel:08}.
A characteristic conical diffraction pattern is shown in Fig. \ref{real_space}(b), 
where the first ring is separated from the second ring that is just starting to 
form (for two distinct rings see \cite{Bahat-Treidel:08}). Experimentally, the 
input beam can be constructed by means of a spatial light modulator (SLM), or by 
three plane waves with wave-vectors corresponding to the three equivalent corners (say $K_+$) 
of the BZ \cite{Bahat-Treidel:08}. 

At high intensities, the nonlinear term comes into play. 
{The input wave-packet has an envelope that possesses circular symmetry. 
Hence, the optical potential induced through the nonlinear term has circular 
symmetry as well. The initial excitation is around the Dirac points, where the 
Dirac cone approximation is excellent. 
Thus, one could} expect that, even though some qualitative differences would 
emerge, the resulting diffraction pattern will still have circular symmetry. 
This educated guess is supported by analysis of the "nonlinear Dirac equation" 
(Dirac equation with a nonlinear term) \cite{PhysRevA.79.053830}. We test this 
idea by solving Eq.(\ref{nls}) numerically using the split-step Fourier algorithm 
in the presence of a focusing nonlinearity, where the input beam is 
comprised of Bloch modes from the vicinity of the point $K_+$.
We construct the input beam using $\bk\cdot\bp$ approximation
\beq
	\psi(\br,0) = \sqrt{\N} [B_{K_+,1}(\br) + i B_{K_+,2}(\br) ] \cdot \exp{\lt(-\br^2/\sg^2\rt)}, 
\eeq
where $B_{K_+,1}(\br)$ and $B_{K_+,2}(\br)$ are the two degenerate Bloch modes 
corresponding to the $K_+$ point, $\sg$ is the width of the envelope, 
and $\N$ determines the beam's power. Surprisingly, we find that the 
diffraction pattern contradicts these expectations by having a three-fold symmetry 
(Fig. \ref{real_space}(c)): instead of the circular rings we obtain a triangle 
pointing to the right. Even more surprising is the diffraction pattern obtained 
when the sign of the nonlinearity is reversed: defocusing instead of focusing 
(Fig. \ref{real_space}(d)). In the defocusing case, the emerging triangular diffraction 
pattern is almost identical to the focusing case, but is rotated by $\pi$. 
The values of the parameters used in the simulation are $D = 35 [\mu m],~\max\{\dl n\} = 7\cdot 10^{-4}
, ~\max\{n_2 |\psi|^2 \} = 5\cdot 10^{-5},~n_0 = 1.5, ~k = 7.85\cdot10^{6}[m^{-1}],$ and the propagation distance $Z = 8 [cm]$,.

We find another interesting phenomenon by repeating the numerical experiment with 
similar input beam that is centered around the other non-equivalent Dirac point 
$K_-$: the orientation of the triangle is reversed again. In other words, for 
a focusing nonlinearity, similar input beams centered around the different 
Dirac points evolve into triangles that are mirror image of each other. 
We emphasize that the $\pi$ rotation of the intensity pattern that results from 
the different Dirac point is {\it exact}, as opposed to the $\pi$ rotation 
resulting from reversing the sign of the nonlinearity. In the latter case, the orientations 
of the triangles are indeed rotated, but the intensities are not identical, as can be 
seen from Fig. \ref{real_space} (c) and (d). Therefore, even though similar 
physical effects are obtained, their physical origin is different.

In all these simulations, we find that the width of the "triangular ring" is {\it smaller}
than the width of the circular ring (arising in the linear case), for both types (signs) of nonlinearity. This 
result is surprising since generally the focusing and the defocusing nonlinearities have 
opposite effects on the width of propagating wave-packet. Namely, in a homogeneous medium, 
the former will always decrease the width of a finite beam, whereas the latter will increase it. These 
opposite tendencies occur also in periodic structures, where the action on the beam width 
is determined by sign of the nonlinearity with respect to the sign of the effective mass, 
e.g., in regions of anomalous diffraction (negative effective mass) a focusing nonlinearity always broadens 
the beam, whereas defocusing narrows it, etc. As stated above, the honeycomb lattice is also 
unique in the sense that the expansion of the evolving triangular ring-beam is not affected 
by the sign of nonlinearity; the width of the ring decreases for both cases (Fig. \ref{real_space}(c) and \ref{real_space}(d)). 
Moreover, the mean radius  
of the triangular ring beam, defined as
\beq
	R \equiv \sqrt{ \bra{\psi(z)}(x^2+y^2)\ket{\psi(z)} },
\eeq
is the same for the linear and nonlinear diffraction patterns, i.e., the 
mean radius of the evolving beam is not affected by the {weak} nonlinearity at all. 
The results of the numerical experiment can be summarized as follows.
\ben
	\item The nonlinearity transforms conical diffraction into triangular diffraction.
	\item The wave-packet evolves into a triangular ring for both focusing and defocusing nonlinearity, and 			nonlinearities of opposite signs result in similar triangles pointing to opposite directions.
	\item Two identical wave-packets centered around different Dirac points yet subject to the same nonlinearity, 			evolve to identical triangular rings pointing in opposite directions.
\een

The remaining part of the paper presents our explanation of these unexpected 
observations that cannot be explained simply by adding a nonlinear term to the Dirac equation.

In order to analyze the dynamics, we write the wave-function as a linear combination of Bloch modes (Bloch decomposition)
\beq
	\psi(\br,z) = \sum_{m} \intl{BZ}{} d^2q B_{\bq,m}(\br) g_{m}(\bq,z), \label{psi expantion}
\eeq
where $g_{m}(\bq,z)$ is the the amplitude of each Bloch mode, and $B_{\bq,m}(\br)$ is { a Bloch wave}
defined in Eq.(\ref{bloch wave}). Substituting the expansion of $\psi$ into Eq.(\ref{nls})  
it is clear that the nonlinear term mixes four Bloch waves with different lattice 
{momenta}, $\bq$, i.e., the nonlinear term generates spatial four-wave-mixing and therefore 
the output beam may include of Bloch modes that were not initially populated. 
Hence, even though the input beam is very localized in $k$-space around the 
Dirac point, it does not necessarily remain localized during the propagation. 
In fact, previous theoretical and experimental work in nonlinear photonic 
lattices has clearly showed that the Bloch population grows by virtue of 
four-wave-mixing \cite{Manela:06}, and in some cases it even leads to spatial supercontinuum \cite{PhysRevLett.101.183903}. 

\begin{figure*}[t]
    \center
    {
    \includegraphics[width=0.96\textwidth]{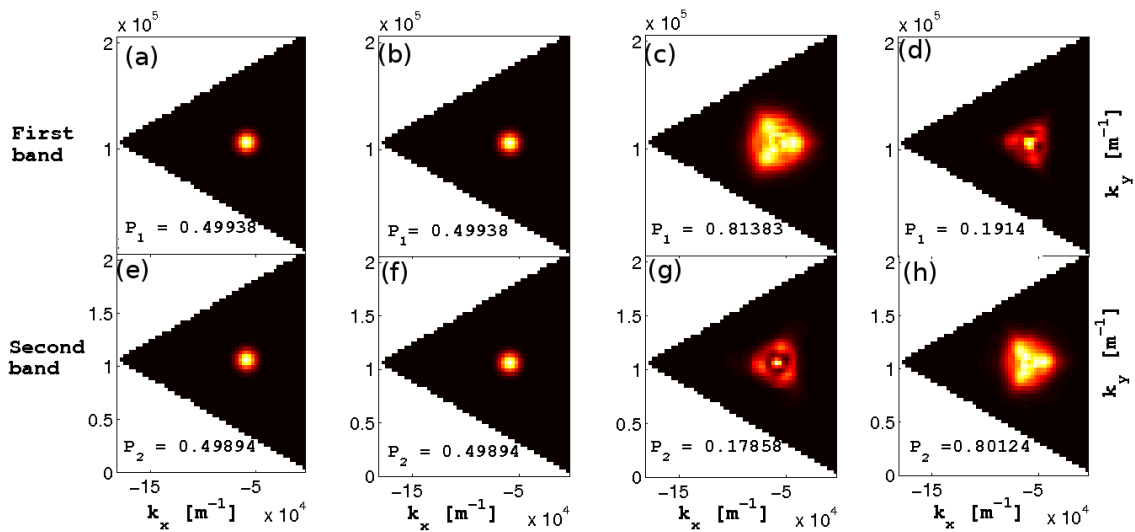}
    }%
    \caption {The bloch mode decomposition of the input wave-packet from the vicinity
     of the Dirac point. (a)-(d) The Bloch distribution of the first band $|g_1(\bq)|^2$,
     and (e)-(h) the Bloch distribution of the second band $|g_2(\bq)|^2$. In each figure we write the 
     corresponding population. Clearly, focusing nonlinearity transfers the energy to the 
     first band, and defocusing non-linearity transfers the population to the second band. }
    \label{bloch_space}
\end{figure*}

\begin{figure}[]
    \center
    {\includegraphics[width=0.96\textwidth]{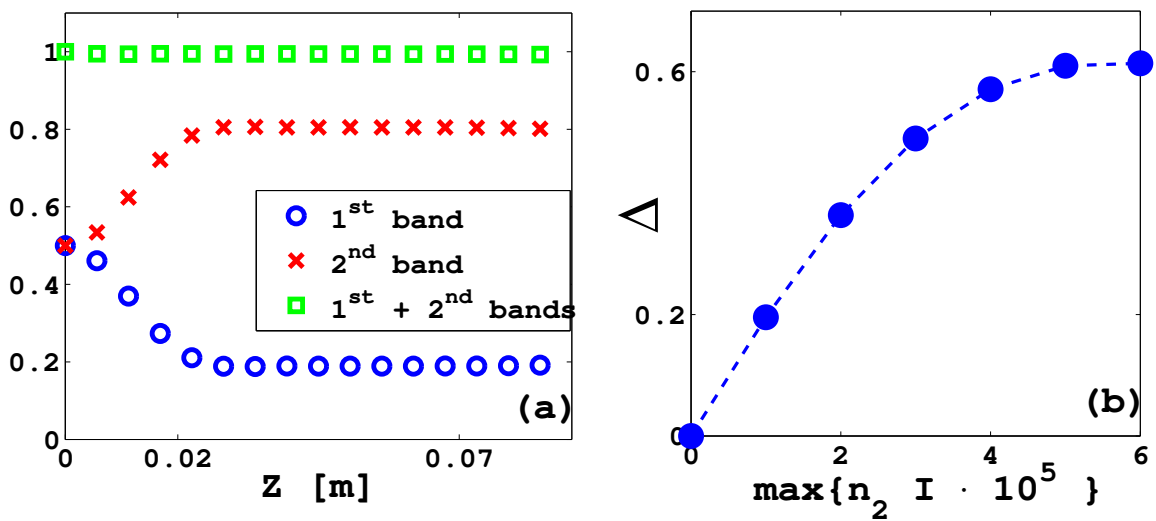}
    }%
    \caption {(a) The population in the first two bands during the propagation in the presence of
    defocusing nonlinearity, where $\max\{|n_2| |\psi|^2 \} = 4\cdot 10^{-5}$. (b) The final population imbalance 
    as a function of the strength of the nonlinear refractive index.}
    \label{population}
\end{figure}

More quantitative analysis of the nonlinear evolution of the distribution of Bloch modes reveals the underlying physics of the triangular diffraction pattern. The population of the $m^{th}$  band is defined as
\beq
	P_m \equiv  \intl{BZ}{} d^2 q |\bracket{\psi(z)}{B_{\bq,m}} |^2 = \intl{BZ}{} d^2q |g_{m}(\bq,z)|^2,
\eeq
and the population imbalance between the first two bands 
\beq
	\Dl(z) = P_1 - P_2 = \intl{BZ}{} d^2q \lt(|g_{1}(\bq,z)|^2 - |g_{2}(\bq,z)|^2 \rt).
\eeq
We calculate the projections of the input and output beams on the Bloch waves of the first 
two bands and find very interesting results: the Bloch distribution $|g_{m}(\bq,0)|^2$ of 
the initial beam has circular symmetry {(as expected)}, with no population imbalance,  
whereas the Bloch distribution of the output beam $|g_{m}(\bq,L)|^2$ has {\bf three-fold symmetry} 
with a significant population imbalance. Fig. \ref{bloch_space} presents the evolution of the 
Bloch distribution for the beam of Fig. \ref{real_space} and the population in each band, 
for both types of nonlinearity. Clearly, focusing nonlinearity increase the population in 
the first band at the expense of the second band, thereby generating a positive imbalance, 
whereas defocusing nonlinearity generates a negative imbalance. While calculating the imbalance, 
$\Dl(z)$, during the propagation we find that it is generated only in the early stages of 
the propagation (Fig. \ref{population}(a)). Moreover, the final imbalance increases with the power of the beam 
(Fig. \ref{population}(b)). 

This analysis leads to three main conclusions: 
\ben
	\item The nonlinearity gives rise to population transfer to Bloch modes that reside outside the Dirac cone, hence, 			{\bf nonlinearity breaks the Dirac dynamics.}
	\item The sign of nonlinearity determines the direction of population transfer.
	\item The Bloch distribution of the output beam always has three-fold symmetry in $k$-space, hence it 
	is clear that the intensity in real space has three-fold symmetry as well.
\een

\begin{figure}[]
    \center
    {\includegraphics[width=0.96\textwidth]{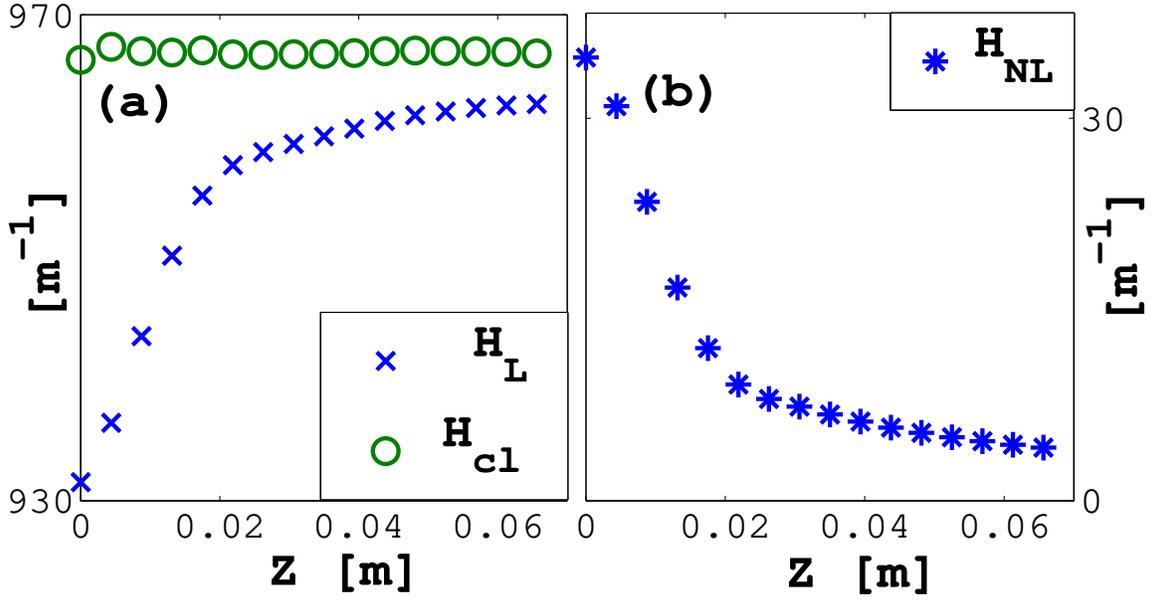}
    }%
    \caption {(a) Calculation of $\H_{cl}(z)$ and $\H_{L}(z)$ in the presence focusing nonlinearity. 
    (b) The nonlinear contribution to $\H_{cl}(z)$, 
    that decreases rapidly during the propagation.}
    \label{beta_z}
\end{figure}

These conclusions can be explained by more fundamental arguments, where the most 
crucial step is to understand why the nonlinearity generates a population 
imbalance between the bands. In order to do so, 
{consider the functional}
\beq
	\H_{cl}(z) = \H_{L} + \H_{NL}, \label{H_cl}
\eeq
where
\beqa
	\H_{L} &\equiv &  -\int \lt(\frac{1}{2k}\lt|\nb_{\perp}\psi\rt|^2 
	- \frac{k}{n_0}\dl n(x,y) |\psi|^2  \rt)d^2x, \quad \mbox{ and}\\
	\H_{NL} &\equiv &
	\int \frac{k}{2n_0}n_2 \lt( |\psi |^2\rt)^2 d^2x.
\eeqa
Note that $\H_{cl}$ is a constant of motion, i.e., $\pad_z \H_{cl} = 0$. 
The evolution equation, Eq.(\ref{nls}), is derived from $\H_{cl}$ via the variational principle 
\beq
	i \pd{\psi}{z} = - \frac{\dl \H_{cl}}{\dl \psi^*}.
\eeq
{It is instructive to note that $\H_{L}$ can be interpreted as the mean} 
propagation constant of the wave-packet in the absence of nonlinearity 
\beq
	\H_{L}(z) = \bra{\psi(z)} \hat H_0 \ket{\psi(z)} = \sum_{m = 1,2} \intl{BZ}{}d^2q
	\beta_{m}(\bq) |g_m(\bq,z)|^2.\label{HL_a}
\eeq
Since $\H_{cl}$ is a constant of motion, we can write
\beq
	\H_{L}(0) + \H_{NL}(0) = \H_{L}(z) + \H_{NL}(z). \label{conservation fo H_cl}
\eeq
The initial beam is constructed around the Dirac point with equal population of both bands. 
Since the bands are almost symmetric, {from Eq.(\ref{HL_a}) we find that
$\H_{L}(0)$ equals} 
the propagation constant at the Dirac point, $\beta_D$. Since the spectrum can be 
shifted by a constant without affecting the dynamics, we can set 
{$\H_{L}(0)=\beta_D=0$}. 
{During} the propagation the beam {experiences} significant 
broadening, and since the total power is conserved the amplitude of the wave-packet 
decreases with $z$. Hence, the nonlinear contribution to $\H_{cl}$ becomes negligible 
compared to the linear contribution at large $z$, i.e., $\H_{NL}(z) \ll \H_{L}(z)$. 
Therefore Eq.(\ref{conservation fo H_cl}) simplifies to
\beq
	\H_{L}(z) \simeq \H_{NL}(0) = \int \frac{k}{2n_0}n_2 \lt( |\psi |^2\rt)^2 d^2x, \label{HL_b}
\eeq
thus the sign of $\H_L(z)$ is identical to the sign of $n_2$.
By substituting Eq.(\ref{HL_a}) and using the symmetry of the bands, $\beta_2(\bq) \approx -\beta_1(\bq)$, we obtain
\beq
	\H_{NL}(0) \simeq \intl{BZ}{}d^2q \beta_{1}(\bq) \lt(|g_1(\bq,z)|^2 - |g_2(\bq,z)|^2\rt). \label{HNL}
\eeq
Since $\beta_1(\bq)$ is positive (first band), the sign of 
$\H_{NL}(0)$ is the same as the final population imbalance, hence the non-linearity determines the direction 
of population transfer between the bands. From Eqs. (\ref{HL_b}) and (\ref{HNL}) it is clear 
that positive $n_2$ (focusing nonlinearity) yields a positive population imbalance, i.e., 
after a fairly large propagation distance, the first band has greater population. 
For both signs of nonlinearities the resulting Bloch distributions are very similar. 
However, since the group velocity of the bands differs in sign, the orientations of the 
resulting triangles are rotated by $\pi$ with respect to each other.
We emphasize that, other than creating a population imbalance and broadening 
the Bloch distribution, the different signs of nonlinearity have the same physical effect, as opposed to other systems.
The arguments given above are supported by numerical calculation of $\H_L, \H_{NL}$ during 
the propagation, as presented in Fig. \ref{beta_z}. Notice that most of the 
population transfer occurs during the early stages of propagation - just as the population imbalance shown in 
{Fig. \ref{population}}(a), and with the same characteristic length.

\begin{figure}[]
    \center
    {\includegraphics[width=0.96\textwidth]{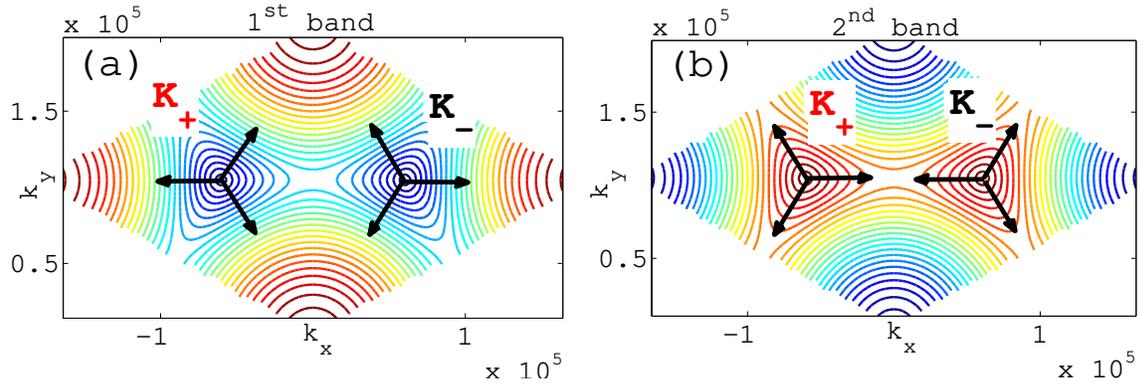}
    }%
    \caption {Contour plot of the first two bands. The Dirac point are denoted by $K_{\pm}$. The black arrows 
    point in the directions where the group velocity is maximal.}
    \label{band_contour}
\end{figure}

Now that we understand that the distribution of Bloch modes varies during propagation, we 
turn to understand the resulting three-fold symmetry. We reexamine the bands around 
the Dirac point by plotting a contour of $\beta(\bq)$ as shown in Fig. \ref{band_contour}, 
and find that very close to the Dirac point, $\beta(\bq)$ has circular symmetry, 
but further away it has three-fold symmetry. Since the new distribution of the Bloch 
modes follows $\beta(\bq)$, the resulting distribution also has three-fold symmetry. 
This argument explains the Bloch distribution of the output beam implying that the 
intensity in real space also has a three-fold symmetry. Due to $z$-reversal symmetry, $\beta(\bq) = \beta(-\bq)$ implying that the contours of equal $\beta(\bq)$ around the point $K_+$ are mirror image of the contour lines around 
the point $K_-$. Therefore, the Bloch 
distribution and hence the real space intensity resulting from the different Dirac 
cones are exact mirror image of each other. 

Next, we try to give more intuitive explanation for the spatial intensity pattern, 
based on the transverse group velocity defined as $\bs v_g \equiv \bs \nb_{\bq} \beta(\bq)$, 
which is in fact the angle of propagation. The input beam populates a region in $k$-space 
in which $\beta(\bq)$ has circular symmetry, and all waves propagate with a 
transverse group velocity of the same magnitude in the radial direction. 
Since the nonlinearity (say, focusing, where we have shown that the nonlinear dynamics is determined by 
the structure of the first band) causes the Bloch distribution to expand in 
$k$-space and leave the Dirac cone, some of the modes propagate with a 
greater group velocity than others. After some propagation distance, these 
modes are mapped in real space to the most distant points which are the vertices 
of the triangle in real space. From Fig. \ref{band_contour} it is clear that around 
the different Dirac points the directions of maximal group velocity are
opposite. Therefore, a beam centered around $K_+$ with focusing 
nonlinearity evolves into a triangle “pointing to the right”, 
whereas the similar input beam centered around $K_-$ with focusing nonlinearity evolves 
into a triangle “pointing to the left” (Fig. \ref{band_contour}(a)). This explains why similar input 
beams around the different Dirac points subjected to the same nonlinearity 
evolve into triangles pointing in opposite directions. 

When the same input beam is subjected to nonlinearity of opposite sign, the 
wave-packet expands in $k$-space as well. However, for the opposite sign of nonlinearity, 
the energy is transferred to the opposite band (as explained earlier). The contour 
lines of both bands are almost identical, but the group velocity has an opposite sign 
(Fig. \ref{band_contour}), and therefore the corners of the intensity triangle (in real space) 
are again opposite to each other. Therefore, an input beam centered around $K_{+} (K_{-})$ 
with {\bf focusing} non-linearity evolves into a triangular ring-beam pointing to the left (right), 
whereas when the same input beam is subjected to a {\bf defocusing} nonlinearity it evolves 
into a triangle pointing to the right (left).

Up to this point the break-down of Dirac dynamics was considered in the context of optics. 
However, we predict the existence of identical phenomena in the context of ultra-cold 
atomic BECs, based on the fact that Eq. (\ref{nls}) describes dynamics of an interacting 
BEC in a honeycomb potential. Such potentials can be constructed optically, however, 
they were considered mainly in the context of fermionic gases (e.g., see Refs. 
\cite{PhysRevLett.99.070401, zhu-2007-98, PhysRevB.77.235107, PhysRevA.80.043411}), 
probably because electrons in real graphene are fermions, for which the ground state 
is a filled Fermi sea up the Fermi level (which can be shifted from Dirac point 
depending on doping), and because the ground state of a Bose gas in a potential 
with graphene band structure is not a condensate occupying Bloch modes in the 
vicinity of the Dirac point. Thus, in order to obtain such a state one should 
do it dynamically. Here we propose a scheme which is not trivial, but still realistic 
experimentally: Suppose that one prepares a finite size condensate in a harmonic, 
pancake shaped, effectively two-dimensional trap \cite{hadzibabic-2009}. At some point 
in time, the in-plane confinement is turned off (the confinement keeping the 2D 
structure is still present), the pancake BEC is subsequently given a momentum kick 
with three plane waves with angles $2\pi/3$ with respect to each other, which are 
directed into the three equivalent corners of the BZ (say $K_+$), and that the 
honeycomb potential is ramped up (the lattice spacing of the honeycomb potential 
should be much smaller than the size of the cloud); the initial BEC can be 
noninteracting, and the nonlinearity can be turned on via the Feshbach mechanism 
(say when the honeycomb potential is ramped up). Such an experimental sequence 
of events is not an easy task but it is within today's experimental capabilities. 
The sign of the nonlinearity and the strength of the interactions can be tuned for some BECs 
by a Feshbach resonance mechanism. 

In conclusion, honeycomb lattices possess intersection points between the first two bands, and Bloch modes that reside close to these points give rise to extremely unique dynamics: dynamics of massless spin half particles described by the Dirac equation. We have shown that, in the presence of nonlinearity, this unique dynamics is significantly altered, and the Dirac equation is no longer suitable for describing the (propagation) evolution. In addition, we studied the nonlinear evolution of wave-packets comprised of such Bloch modes, which under linear conditions yield conical diffraction. We have found that both signs of nonlinearity transform the circular rings into triangular rings, resulting in “triangular diffraction pattern”. This new type of diffraction cannot be obtained simply by adding the nonlinear term into the effective Dirac equation, since the Bloch modes distribution quickly expands beyond the range of the Dirac cone, where the effective evolution equation governing the dynamics cannot be approximated by the Dirac equation, nor by its nonlinear extension. Hence we expect that the intriguing phenomena obtained in honeycomb lattices such as Klein tunneling, Zitterbewegung, anti-localization, zero modes edge-states \cite{PhysRevB.54.17954, PhysRevB.76.205402} and more would be significantly altered by the presence of nonlinearity. Likewise, the intriguing features related to Dirac points in photonic crystals \cite{PhysRevLett.100.013904, PhysRevA.75.063813, PhysRevB.78.045122, PhysRevB.80.155103} would be affected by nonlinearities of the same type studied here.

\acknowledgments

This work is supported in part by the Croatian-Israeli scientific collaboration funded by the ministries of science 
in both countries, and by an advanced grant from the ERC.
H.B. acknowledges support from Croatian Ministry of Science (Grant No. 119-0000000-1015).

\end{document}